\documentclass[aps,prb,nofootinbib,twocolumn,amsmath,amssymb,superscriptaddress,notitlepage]{revtex4-2}
\usepackage{latexsym}
\usepackage{graphicx}
\usepackage{times,psfrag,subfigure}
\usepackage{enumerate}

\usepackage{mathtools}
\usepackage{amsthm}
\usepackage{dsfont}
\usepackage{dcolumn}
\usepackage{bm,bbm}
\usepackage{physics}

\usepackage{wasysym}
\usepackage{tikz-cd}
\usepackage{tikz}
\usetikzlibrary{calc}
\usepackage[normalem]{ulem} 
\usepackage{color}
\usepackage[dvipsnames]{xcolor}
\usepackage[breaklinks,colorlinks,allcolors=blue]{hyperref}
\usepackage{latexsym,amsmath,amssymb,bm,euscript}
\usepackage{textcomp}
\usepackage{tabularx}
\usepackage{multirow}
\usepackage{setspace}
\usepackage{ctable}
\usepackage{sidecap}
\usepackage{placeins}
\usepackage{threeparttable}
\usepackage{braket}
\usepackage{multirow}
\usepackage{comment}
\usepackage{float}
\usepackage{framed}
\usepackage{scalerel,stackengine}
\stackMath

\usepackage{ulem}

\def\S{\mathcal{S}}

\def\bbD{\mathbb{D}}
\def\bbZ{\mathbb{Z}}

\newcommand\reallywidehat[1]{%
\savestack{\tmpbox}{\stretchto{%
  \scaleto{%
    \scalerel*[\widthof{\ensuremath{#1}}]{\kern-.6pt\bigwedge\kern-.6pt}%
    {\rule[-\textheight/2]{1ex}{\textheight}}
  }{\textheight}%
}{0.5ex}}%
\stackon[1pt]{#1}{\tmpbox}%
}

\makeatletter
\renewcommand*\env@matrix[1][*\c@MaxMatrixCols c]{%
  \hskip -\arraycolsep
  \let\@ifnextchar\new@ifnextchar
  \array{#1}}
  
\makeatother

\allowdisplaybreaks

\graphicspath{{figures/}}
\begin{document}

\title{On Lagrangians of Non-abelian Dijkgraaf-Witten Theories}

\author{Yuan Xue}
\affiliation{Department of Physics, The University of Texas at Austin, Austin, Texas 78712, USA}

\author{Eric Y. Yang}
\email{yuy073@ucsd.edu}
\affiliation{Department of Physics, University of California, San Diego, 9500 Gilman Drive \# 0319
, La Jolla, CA 92093, USA}

\date{\today}

\begin{abstract}
    Dijkgraaf-Witten theories have a wide range of applications in topological phases of matter and the study of generalized global symmetries. We develop a method to construct BF-type Lagrangians for Dijkgraaf-Witten theories with non-abelian gauge group by gauging $H^{(0)}$ symmetries from a BF-Lagrangian of an abelian Dijkgraaf-Witten theory. When $H$ nontrivially permutes the operators of the original theory, the Lagrangian of the $H$-gauged theory is constructed with cohomologies with local coefficients. We analyze the structure of the Lagrangians and their gauge transformations with homotopy theory. We also construct the operator spectrum and verify the Lagrangians by matching elementary linking invariants.
\end{abstract}

\maketitle

\tableofcontents

\section{Introduction}

Discrete gauge theories \cite{BANKS198990,wilzeck,PRESKILL199050} have a wide range of applications in condensed matter physics and high energy physics. An important class of examples are the topological discrete gauge theories as discussed by Dijkgraaf and Witten \cite{Dijkgraaf:1989pz} in $D=3$. We will here refer to such TQFT discrete gauge theories in $D\geq 3$ as ``Dijkgraaf Witten” (DW) theories. In condensed matter physics, DW theories describe the long-wavelength limit of a wide class of $D$-dimensional bosonic topological orders for $D\geq 3$ \cite{PhysRevB.71.045110,PhysRevB.87.155114,PhysRevB.82.155138,doi:10.1142/S0217979290000139,PhysRevX.8.021074}, where $D$ is the spacetime dimension. When the spacetime manifold $M_D$ is given a triangulation, one can explicitly define the partition function of the DW theory. There also exist lattice Hamiltonian constructions of DW theories in (2+1)D \cite{Kitaev:1997wr,Hu:2012wx} and (3+1)D \cite{Wan:2014woa}. Recently, DW theories have also regained popularity in high-energy physics due to the modern perspective of generalized symmetries \cite{Gaiotto:2014kfa}. The central paradigm is that global symmetries of a quantum field theory, topological or not, are implemented by topological defects. The topological nature of the global symmetries of a theory $\mathcal{T}$ on $M_D$ motivates the definition of the symmetry topological field theory (SymTFT) \cite{Freed:2022qnc} on a cylinder $M_D\times I$, which captures the symmetry data, defects, and anomalies of $\mathcal{T}$ as topological operators and boundary conditions. See \cite{Bhardwaj:2023kri,Schafer-Nameki:2023jdn} and the references therein for further details. The same construction also appears in the condensed matter theory literature as SymTO/topological holography \cite{Chatterjee:2022kxb, Chatterjee:2022tyg}.

Mathematically, DW theories are extended TQFTs \cite{Baez:1995xq,Lurie2009OnTC}, and a proper mathematical investigation would require higher category theory. Given its relevance in both the condensed matter theory and the high energy theory community, it is helpful to formulate a purely field theoretic treatment of DW theory where operator fusion and linking can be studied from elementary operator algebra and path integral arguments familiar to physicists. If a DW theory contains only invertible operators, then in many cases it can be formulated as a BF theory \cite{Kapustin:2014gua}. Meanwhile, abelian DW theories with certain non-trivial DW twists can contain non-invertible operators. Lagrangian analysis of such theories has previously appeared in the context of linking invariant calculations in non-abelian topological orders \cite{Putrov:2016qdo, He:2016xpi,Gao:2025ihw}, example studies of symTFT for categorical symmetries \cite{Kaidi:2023maf,Robbins:2025puq}, as well as top-down derivation of symTFT from IIA/IIB/F-theory \cite{Yu:2023nyn, Franco:2024mxa,Bergman:2024its, Heckman:2022xgu}. However, a universal bottom-up Lagrangian construction with a non-abelian gauge group $G$ is still lacking. 

This work aims to provide a Lagrangian treatment of DW theories with non-abelian gauge group $G$ that fits into an abelian extension:
\begin{equation}
    0 \rightarrow A \rightarrow G \rightarrow H \rightarrow 0
\end{equation}
where $A$ and $H$ are both abelian. We would like to construct a BF-type Lagrangian for a $G$-DW theory from the BF-type Lagrangian of an $A$-DW theory. Coincidentally, the same construction also appeared in the previous studies of finite 0-form symmetry gaugings in DW theories \cite{Kapustin:2014zva}, where the extension class $c_2\in H^2(BH,A)$ encodes the fractionalization of the $H$ 0-form symmetry. In this work, we will construct $G$-DW theory Lagrangians by gauging an $H$ 0-form symmetry in an $A$-DW theory. When $H$ nontrivially acts on the group $A$, the new gauge group $G$ is necessarily non-abelian, and the topological action of the resulting $G$-DW theory is described in terms of cohomologies with local coefficients. Partial progress has been made in our previous work \cite{Xue:2025enx}, where we proposed a mechanism for constructing $G$-Lagrangians from $A$-Lagrangians by gauging global symmetries via an effective Noether procedure. The Poincar\'e dual of the Noether currents are represented by higher gauging condensation defects \cite{Roumpedakis:2022aik}, whose classification in untwisted DW theories was examined in \cite{Cordova:2024jlk, Cordova:2024mqg}. As in our previous work, the guiding principle of the operator construction is operator world-volume gauge invariance, and we verify our constructions by showing that the linking invariant calculations can correctly reproduce the character table of $G$. 

This paper is summarized as follows. In Sec. \ref{section-Lagrangians}, we will give a brief review of DW theories and present the recipe for constructing the $G$-action. For simplicity, we take $G$ to be the dihedral group of order $2k$ $\mathbb{D}_k\simeq \mathbb{Z}_k \rtimes \mathbb{Z}_2$ in $(3+1)$D. In this case, the twist coincides with a $\mathbb{Z}_2^C$ 0-form charge conjugation symmetry of the original $\mathbb{Z}_k$-DW theory without fractionalization. We will also carefully study the gauge transformation structure and explain its origin in terms of homotopy theory assuming a knowledge of the theory of smooth fiber bundles.  We will postpone the technical homotopy-theoretic details to Appendix \ref{appendix-homotopy}. When gauging a finite symmetry in any theory, one necessarily needs to verify that the symmetry action is anomaly-free. We will address this in Sec. \ref{section-anomalies}. In Sec. \ref{section-linkings}, we will verify our Lagrangian construction by demonstrating how to extract the character table of $G$ from Hopf links between Wilson lines and 't Hooft surfaces and tabulate the full list of results in Appendix \ref{appendix-linking-results}. In Sec. \ref{section-conclusion}, we will summarize and point out possible future directions.

\bigskip

\textit{\textbf{Notes added}}: During the preparation of this draft, we became aware of \cite{Bergman:2026lnz}, which partially overlaps with our results. \cite{Bergman:2026lnz} decomposes a $G$-action in terms of $A$-gauge fields and $H$-gauge fields and the Lagrangian examples are mostly in $(2+1)D$. This work focuses on gauging a $H$-symmetry in an $A$-action in $(3+1)$D and has a new family of examples for $G=\mathbb{D}_{k}$, where $\mathbb{D}_{k}$ is the dihedral group of order $2k$ for $k$-even. Our organization of the operator spectrum can be easily generalized to any spacetime dimension $D\geq 3$. 

\section{Lagrangian Constructions and Gauge Transformations}\label{section-Lagrangians}

In this section, we construct the Lagrangians of $\mathbb{D}_{k}$-DW theories by gauging the $\mathbb{Z}_2^C$ symmetry in an untwisted $\mathbb{Z}_k$-DW theory. In the spirit of \cite{Kapustin:2014zva}, these are candidate actions of some 0-form symmetry gauged actions that are possibly obstructed by 't Hooft anomalies. We will postpone the 't Hooft anomaly analysis to Sec. \ref{section-anomalies}. Central to our analysis is the typical BF-theory formulation of DW theory, which we will quickly review in Sec. \ref{BF-theory-formulation}. In Sec. \ref{charge-conjugation-gauged-action}, we will construct the BF theory action of untwisted $\mathbb{D}_{k}$-DW theories. Note that the group $\mathbb{D}_4$ is rather special, as it fits in various central and split extensions. We will construct various versions of its Lagrangian in Sec. \ref{D4-action}.

\subsection{BF Theory Formulation of Discrete Gauge Theories} \label{BF-theory-formulation}

In this subsection, we quickly review the definition of DW theories and their BF-theory formulations. Instead of using the extended TQFT definition in the framework of \cite{Baez:1995xq,Lurie2009OnTC}, it is convenient to treat DW theories as an analog of Yang-Mills theories with a finite gauge group $G$ \cite{Dijkgraaf:1989pz}. In this subsection, we set the spacetime manifold to be a closed oriented $D$-dimensional manifold $M_D$ with a triangulation. 

Just like ordinary Yang-Mills theory with a compact continuous gauge group, the structure of a DW theory can be described by principal $G$ bundles $P \rightarrow M_D$. A ``classical" DW theory is specified by a choice of gauge group $G$ and a topological action $\omega\in H^D(BG,U(1))$, which can be interpreted as the universal cocycle data of a $G$-SPT on $M_D$ \cite{Chen:2011pg}. When $\omega$ is the identity element, the theory is typically referred to as an untwisted $G$-DW theory. Here $BG$ is the classifying space of $G$, which is a topological space whose only non-vanishing homotopy group is the fundamental group $\pi_1(BG) = G$. Quantization of the theory is physically described by gauging the $G$-symmetry in the $G$-SPT, which results in a nontrivial topological order with loop operators/ Wilson lines and codimension-2 surface operators/ 't Hooft surfaces in spacetime. When the topological action $\omega$ is trivial, the line operators are in one-to-one correspondences with the linear irreps of $G$, and the codimension-2 surface operators are in one-to-one correspondences with the conjugacy classes of $G$.

Topologically, quantization is described by defining the partition function as a weighted sum over isomorphism classes of $G$-bundles over $M_D$. It is well known that these bundles are classified by the classifying maps $f: M_D \rightarrow BG$ up to homotopy. Specifically, for any $BG$ there exists a contractible space $EG$ called the universal covering space which admits a free $G$ action and the quotient $EG/G$ is homotopic to $BG$. It is then a standard math fact that any principal $G$ bundle $P\rightarrow M_D$ is isomorphic to the pull-back bundle induced by the classifying map $f: M_D \rightarrow BG$:
\begin{equation}
    \begin{tikzcd}
P = f^*(EG) \arrow[r,"f^*"] \arrow[d] & EG \arrow[d] \\
M_D \arrow[r,"f"] & BG
\end{tikzcd}
\end{equation}
The partition function is given by:
\begin{equation}
Z_{\omega}(M_D)
=
\frac{1}{|G|}
\sum_{\phi \in \mathrm{Hom}(\pi_1(M_D),G)}
\left\langle \phi^{*}\omega,[M_D]\right\rangle.
\end{equation}
After modding out gauge transformations on the path integral measure, using the isomorphism $[M_D, BG] \simeq \mathrm{Hom}(\pi_1(M_D),G)/G$\footnote{It is a highly nontrivial fact that $[M_D, BA]$ has the structure of a finite abelian group for a finite abelian $A$. A full explanation involves the definition of the based loop space $\Omega X$ of $X$. We refer the readers to \cite{HatcherAT} for further details.} where $[M_D, BG]$ denotes the homotopy classes of maps from $M_D$ to $BG$,  one can rewrite the partition function as:
\begin{equation}
Z_\omega(M_D)
=
\sum_{[\rho]\in [M_D, BG] }
\frac{1}{|C_G(\rho)|}
\,
\left\langle \rho^*\omega,[M_D]\right\rangle,
\end{equation}
where \(C_G(\rho)\) denotes the centralizer of the image of \(\rho\) in \(G\). Crucially, since the quantization procedure involves $[M_D, BG]$, DW theories are also termed as \textit{topological sigma models} in the literature  \cite{Delcamp:2019fdp}. 

When $G=A$ is abelian, there is an alternative definition in terms of gauge fields. Using the isomorphism $H^n(BA, U(1))\simeq H^{n+1}(BA, \mathbb{Z})$, one can rewrite the inner product $\left\langle \rho^*\omega,[M_D]\right\rangle$ as a phase $\exp(2\pi i\int_{[M_D]}\rho^*\omega)$ and replace the measure with the  $[M_D, BA]\simeq H^1(M_D, A)$. For example, since $H^4(B\mathbb{Z}_k,U(1)) = \mathbb{Z}_1$, $\mathbb{Z}_k$-DW theories in $(3+1)$D can only be defined with respect to a trivial topological action, and its partition function reads:
\begin{equation}
    Z(M_D) = \sum_{[\rho]\in H^1(M_D,\mathbb{Z}_k)}\frac{1}{\abs{C_A(\rho)}} = \frac{\abs{H^1(M_D,\mathbb{Z}_k)}}{\abs{Z_k}}
\end{equation}
When $M_D$ is simply connected, the partition function equals $1/\mathbb{Z}_k$ and when $M_D$ is torsionless, the partition function equals $k^{b_1(M_D)-1}$.

The fact that the path integral measure is a sum over $H^1(M_D,A)$ is the main motivation for the BF-theory formulation. Suppose $A = \mathbb{Z}_k$ for concreteness. A 1-form gauge field $a_1$ is a $\mathbb{Z}_k$-valued 1-cocycle on $M_D$ taking values in $\{0,1,\dots, k-1\}$. In Lagrangian formulations, it is convenient to work with integer lifts of $\mathbb{Z}_k$-valued gauge fields, where $\mathbb{Z}_k$ fits in the following short exact sequence:
\begin{equation}
    0 \rightarrow \mathbb{Z} \xrightarrow{\times k} \mathbb{Z}\xrightarrow{\text{mod }k }\mathbb{Z}_k \rightarrow
     0
\end{equation}
 The BF-theory Lagrangian is defined by introducing a $\widehat{\mathbb{Z}}_k$ Lagrange multiplier $\tilde{b}_{D-2}$ \cite{Kapustin:2014gua}:
\begin{equation} \label{eq-BF-action}
    I = \frac{2\pi}{k} \int_{M_D} \tilde{b}_{D-2}\cup \delta b_1 + \int_{M_D} (\rho^*\omega)(b_1)
\end{equation}
so that the flatness constraint of $b_1$ is encoded in the equation of motion for $\tilde{b}_{D-2}$. Integrating out $\tilde{b}_{D-2}$ recovers the original definition of DW action. BF-type actions are convenient for an explicit operator algebra analysis of the TQFT. On the other hand, the action obtained by integrating out the Lagrange multiplier fields are convenient for the purpose of 't Hooft anomaly analysis \cite{Kapustin:2014zva}. 

Finally, we point out a crucial subtlety of the BF-theory/ Lagrangian formulation of DW theories. Notice that the group homomorphism $f^*:H^D(Y,U(1))\rightarrow H^D(X,U(1))$ induced by the map $f:X \rightarrow Y$ in general can have a nontrivial kernel. In the case of DW theory, we simply replace $X$ with a physical spacetime $M_D$ and $Y$ with the target space $BG$ for any finite group $G$. Adopting the topological sigma model definition, this means that a nontrivial topological action $\omega \in H^D(BG,U(1))$ can be trivialized once pulled back to spacetime if $\ker f^*$ is not trivial, where
\begin{equation}
   f^* : H^D(BG,U(1)) \rightarrow H^D(M_D, U(1))
\end{equation}
Namely, if $\omega_G\in \ker f^*$, then its action on spacetime $f^*\omega_G$ is necessarily trivial. Since certain nontrivial topological actions $\omega_D\in H^D(BG,U(1))$ can be pulled back to the trivial cocycle on spacetime $M_D$, working with a BF-type Lagrangian on spacetime alone does not capture all the data of the DW theory, and the specification of the initial data $\omega_G\in H^D(BG,U(1))$ is necessary.

\subsection{Charge Conjugation Gauged Lagrangian} \label{charge-conjugation-gauged-action}

In this subsection, we construct the BF-Lagrangians of charge conjugation gauged $\mathbb{Z}_k$ DW theories in (3+1)D without symmetry fractionalization and study the structure of their gauge transformations. In particular, we combine the field theoretic on-shell and off-shell deformations and show how they can be reproduced from a canonical homotopy theory perspective \cite{Freed2022FOURLO}.

Let $a_1$ be a $\mathbb{Z}_k$ valued cochain. We can gauge the charge conjugation by a two-step procedure. Since the charge conjugation symmetry non-trivially permutes the extended operators, the coupling of the background gauge fields necessarily changes the cohomological structure of the theory \cite{Kaidi:2022cpf} and modifies the original untwisted action to\footnote{When the 1-cochain $c_1$ is used only to indicate the twisting in the subscript, we will simply write $c$ instead of $c_1$. Similarly, for the gauge transformation $c_1\rightarrow c_1 + \delta \epsilon_0$, we shall only write $\epsilon$.}
\begin{equation}
    I = \frac{2\pi}{k}\int \Tilde{a}_2\cup_c\delta_c a_1
\end{equation}
where $c_1$ is a $\mathbb{Z}_2$-valued 1-cocycle. To complete the gauging, we promote $c_1$ to dynamical gauge fields and allow off-shell fluctuations of the $c_1$ gauge field:
\begin{equation} \label{eq-CC-gauged-Lagrangian}
    I_{\mathbb{D}_k} = \frac{2\pi}{k}\int_{M_4} \Tilde{a}_2\cup_c\delta_c a_1 + \pi\int_{M_4} \Tilde{c}_2\cup \delta c_1
\end{equation}
where $\Tilde{c}_2$ is a $\widehat{\mathbb{Z}}_2$-valued 2-cochain implementing the flatness constraint on $c_1$.

The on-shell and off-shell physics of Eq. \eqref{eq-CC-gauged-Lagrangian} admit an interesting hierarchy structure. Especially, we would like to study the on-shell and off-shell deformations that leave the action Eq. \eqref{eq-CC-gauged-Lagrangian} invariant. For clarity, we note that on-shell deformations should not introduce new solutions or eliminate existing solutions to the equations of motion in a usual field theory sense. On the other hand, this requirement does not apply to off-shell deformations as they need not respect the equations of motion.

Let us first examine the on-shell physics. For simplicity, we integrate out the Lagrange multipliers, which will be later restored. The relevant equations of motion are:
\begin{equation}
    \begin{split}
        \delta_{c} a_1 = 0 \qquad
         \delta c_1 = 0 
    \end{split}
\end{equation}
which implies that $a_1 \in H^1_{c}(M_D,\mathbb{Z}_k) $ 
where $H^\bullet_{c}(M_D,\mathbb{Z}_k)$ is a twisted cohomology theory defined with respect to a twisted coboundary operator $\delta_{c}$. In the language of \cite{Kapustin:2014zva}, the twisted flatness of $a_1$ is a consequence of the vanishing of symmetry fractionalization. 

Now we run into an immediate problem. The flatness constraint $\delta c_1 = 0$ implies that the on-shell physics for $c_1$ is invariant under a deformation by a coboundary $c_1 \mapsto c_1 + \delta \epsilon_0$, and the solution space modulo deformation is $c_1 \in H^1(M_D,\mathbb{Z}_2)$. However, the twisted flatness constraint for $a_1$ implies that each particular closed profile of $c_1$ defines a particular cohomology theory that allows for the definition of the action in Eq. \eqref{eq-CC-gauged-Lagrangian}. If we allow a deformation $c_1\mapsto c_1 + \delta \epsilon_0$, then the twisted coboundary operator $\delta_c$ will be deformed into $\delta_{c+\delta \epsilon}$. This deformation nontrivially maps the equation of motion $\delta_{c} a_1 = 0$ to $\delta_{c+\delta\epsilon}a_1= 0$. Accordingly, we will have a new solution space $H^1_{c+\delta\epsilon}(M_D,\mathbb{Z}_k)$. However, since there is no canonical isomorphism between $H_{c}^\bullet(M_D,\mathbb{Z}_k)$ and $H_{c + \delta \epsilon}^\bullet(M_D,\mathbb{Z}_k)$, it follows that a particular $\delta_c$-cocycle on spacetime $M_D$ might not be a $\delta_{c+\delta\epsilon}$-cocycle.  Therefore, a $c_1 \mapsto c_1 +\delta\epsilon_0$ deformation on-shell does not preserve the on-shell physics. In fact, it does something worse --- it changes the equations of motion themselves. In this sense, it is not a genuine on-shell deformation. Therefore, the allowed on-shell gauge transformation should be:
\begin{equation}\label{eq-on-shell deformation}
 a_1 \mapsto a_1 + \delta_{c}\alpha_0\qquad
    c_1 \mapsto c_1 
\end{equation}

This leads to an interesting puzzle. From a field theory perspective, it is unphysical to demand that the solutions to a particular equation of motion $\delta c_1 = 0$ have no non-trivial deformations. On the other hand, nontrivial deformation on $c_1$ is incompatible with the on-shell physics of $a_1$. Luckily, this puzzle can be resolved by going to the quantum theory, where we are no longer constrained by the equations of motion.

Now we address the off-shell gauge transformations, which need not preserve the equations of motion, but they do need to leave the quantum theory invariant. For DW theories, this means that any off-shell gauge transformation must only deform the pulled back topological action $\rho^*\omega$ by a spacetime coboundary. To fully understand the off-shell gauge transformations, it is convenient to invoke the topological sigma model definition of DW theory \cite{Freed2022FOURLO}, where the gauge transformations can be understood as deformations of maps from the physical spacetime $M_D$ to the target space.

Recall that a DW theory with gauge group $\mathbb{D}_{k}$ is a topological sigma model from $M_D$ to the classifying space $B\mathbb{D}_k$. Since $\mathbb{D}_k$ sits in a split extension:
\begin{equation}\label{eq - extension}
    0 \rightarrow \mathbb{Z}_k \rightarrow \mathbb{Z}_k \rtimes \mathbb{Z}_2 \rightarrow \mathbb{Z}_2 \rightarrow 0
\end{equation}
we can model the target space $B\mathbb{D}_k$ as the total space of a Serre fibration of classifying spaces\footnote{Strictly speaking, only the map $\pi: B\mathbb{D}_k\rightarrow B\mathbb{Z}_2$ is called a Serre fibration, and the entire Eq. \eqref{eq-Serre-fibration} is called a fiber sequence. The topological spaces are presented in such a way that is reminiscent of the short exact sequences of the finite groups that induces this fiber sequence. }:
\begin{equation} \label{eq-Serre-fibration}
    B\mathbb{Z}_k \rightarrow B\mathbb{D}_{k} \xrightarrow{\pi} B\mathbb{Z}_2
\end{equation}
For clarity, let us first consider a simplified setup by replacing the fiber $B\mathbb{Z}_k$ with an empty set, which reduces the target space to $B\mathbb{Z}_2$. The reduced theory has the following standard BF action:
\begin{equation}
    I_{\text{reduced}} = \pi \int_{M_4}\Tilde{c}_2\cup\delta c_1
\end{equation}
with equations of motion $\delta c_1 = 0$ and $\delta \Tilde{c}_2 = 0$. In this case, the on-shell and off-shell small gauge transformations coincide:
\begin{equation}
    c_1 \mapsto c_1 + \delta\epsilon_0 \qquad \Tilde{c}_2 \mapsto \Tilde{c}_2 + \delta \Tilde{\epsilon}_1
\end{equation}
Especially, because of the isomorphism:
\begin{equation}
    H^1(M_4,\mathbb{Z}_2) \simeq [M_4, B\mathbb{Z}_2]
\end{equation}
the small gauge transformation is the cohomological analog of a null-homotopy of a particular map from $M_4$ to $B\mathbb{Z}_2$. 

Now let us turn the homotopy fiber back on and consider the untwisted $\mathbb{D}_k$ DW theory. Here we will give a homotopy-theoretic interpretation of the off-shell gauge transformations of an untwisted $\mathbb{D}_k$ theory\footnote{The fact that one can find homotopy theory analogs of cohomological operation is a consequence of the \textit{representability} of simplicial cohomology. We refer the readers to \cite{HatcherAT} for further details. }. We will see that this picture naturally encodes the gauge transformation $c_1\rightarrow c_1 + \delta\epsilon_0$, which is absent from the on-shell deformations. 

Similar to the case where we trivialized the fiber, the collection of gauge fields for untwisted $\mathbb{D}_k$ DW theories are described in terms of homotopy classes of maps $[M_4, B\mathbb{D}_{k}]$. However, since $\mathbb{D}_k$ is non-abelian, we cannot naively define $\mathbb{D}_k$-valued 1-cocycles on $M_4$. Instead, one can try to use a pair of $\mathbb{Z}_k$ and $\mathbb{Z}_2$ gauge fields to collectively denote a ``$\mathbb{D}_{k}$ gauge field". To find a homotopy theory analog of this decomposition, let's try to find inspirations by recalling certain features of smooth fiber bundles, which are a special class of homotopy fibrations. Recall that the total space of a smooth fiber bundle $E \rightarrow B$ admits a local trivialization so that locally we can always describe the geometry of total space $E$ in terms of the base $B$, the fiber $F$ and a section $s: B\rightarrow E$. Going back to a generic Serre fibration $E\rightarrow B$ with typical fiber $F$, one may ask if it is possible to describe the space $E$ in terms of the space $B$ and the $F$. The answer is somewhat positive, and extra care must be taken. For the readers unfamiliar with homotopy theory, in order to understand the intuition in the main text, it suffices to replace the word ``fibration" with a ``smooth fiber bundle". We will address the homotopy-theoretic details in Appendix \ref{appendix-homotopy}.

Consider any map $f: M_4 \rightarrow B\mathbb{D}_k$, which physically describes a ``mode" integrated over in the path integral measure of the $\mathbb{D}_k$ DW theory. The map $f$ can be trivially composed with the map $\pi: B\mathbb{D}_k \rightarrow B\mathbb{Z}_2$ in Eq. \eqref{eq-Serre-fibration} into $\Tilde{f} \equiv \pi\circ f$ so that we have a commutative triangle:
\begin{equation}
    \begin{tikzcd}
        & B\mathbb{D}_k \arrow[d,"\pi"] \\
        M_4 \arrow[ur,"f"] \arrow[r,"\tilde f"'] & B\mathbb{Z}_2
    \end{tikzcd}
\end{equation}
Namely, for any $x\in M_4$, we have $\tilde{f}(x) = (\pi\circ f) (x)$. The map $\Tilde{f}: M_4 \rightarrow B\mathbb{Z}_2$ is a homotopy theory analog of the $\mathbb{Z}_2$ sector of a ``$\mathbb{D}_k$ gauge field".

To get the analog of the $\mathbb{Z}_k$ sector gauge field, we temporarily suppress the map $f: M_4 \rightarrow B\mathbb{D}_k$ in the above diagram and consider the pull-back fibration by the map $\tilde{f}: M_4 \rightarrow B\mathbb{Z}_2$. Note that a pull-back fibration is simply a homotopy-theoretic analog of a pullback smooth fiber bundle. Similar to a pullback smooth fiber bundle, a pullback fibration has the same typical fiber as the original fibration, so now we can expand this triangular diagram into:
\begin{equation}
    \begin{tikzcd}
        & B\mathbb{Z}_k \arrow[d] \arrow[dl]\\
        \tilde{f}^*B\mathbb{D}_k \arrow[r,dashed] \arrow[d, "\tilde \pi"] & B\mathbb{D}_k \arrow[d,"\pi"] \\
        M_4  \arrow[r,"\tilde f"'] & B\mathbb{Z}_2
    \end{tikzcd}
\end{equation}
One can define a section of the pull-back fibration $B\mathbb{Z}_k \rightarrow \tilde{f}^*B\mathbb{D}_k \xrightarrow{\tilde\pi} M_4$, which is a map $s_{\tilde \pi}: M_4 \rightarrow \tilde{f}^*B\mathbb{D}_k$. The fiber direction of this section is the homotopy-theoretic analog of the $\mathbb{Z}_k$ gauge field. To cleanly illustrate this construction, we restore the original effective ``$\mathbb{D}_k$" gauge field as a blue arrow in the following diagram and identify the decomposition into the $\mathbb{Z}_k$ and $\mathbb{Z}_2$ components with blue arrows:
\begin{equation}\label{eq-homotopy-decomposition}
    \begin{tikzcd}
        & B\mathbb{Z}_k \arrow[d] \arrow[dl]\\
        \tilde{f}^*B\mathbb{D}_k \arrow[r,dashed] \arrow[d, "\tilde \pi"] & B\mathbb{D}_k \arrow[d,"\pi"] \\
        M_4 \arrow[ur,red, "f"]\arrow[u, blue, bend left=35,"s_{\tilde\pi}"]  \arrow[r,blue,"\tilde f"'] & B\mathbb{Z}_2
    \end{tikzcd}
\end{equation}

In this language, both the on-shell and off-shell gauge transformations of the gauge fields are described by deformations of the maps in Eq. \eqref{eq-homotopy-decomposition}. We see that it makes sense to discuss a fiber-wise deformation while holding $\tilde f$ fixed. This corresponds to an on-shell gauge transformation with respect to a fixed $c_1$ profile in Eq. \eqref{eq-on-shell deformation}. In homotopy theory terms, such a deformation is simply a change of section $s_{\tilde\pi}$ of the pull-back fibration. 

Since the path integral measure integrates over all possible maps from $M_4$ to $B\mathbb{Z}_2$ modulo homotopy, the off-shell gauge transformation necessarily includes a shift $c_1 \rightarrow c_1 + \delta \epsilon_0$ as expected. From the homotopy-theoretic perspective, we see that the section $s_{\tilde \pi}: M_4 \rightarrow \tilde f^* B\mathbb{D}_k$ is determined by the map $\tilde{f}: M_4 \rightarrow B\mathbb{Z}_2$. Namely, the $\mathbb{Z}_k$ components of the $\mathbb{D}_k$ can only be defined with respect to a particular $\mathbb{Z}_2$ gauge field. Therefore, a deformation on the map $\tilde f$ necessarily induces a change on the section $s_{\tilde \pi}$. In terms of gauge fields, this corresponds to the following off-shell gauge transformation:
\begin{equation}
    \begin{split}
        c_1 &\mapsto c_1 + \delta \epsilon_0 \\
        a_1 &\mapsto a_1 + \delta_{c+\delta \epsilon}\alpha_0
    \end{split}
\end{equation}

Now we address the off-shell gauge transformations of the Lagrange multiplier fields. Since the $c_1$-twist acts on $\mathbb{Z}_k$ as an automorphism, the off-shell gauge transformation of the $\widehat{\mathbb{Z}}_k$-valued gauge field $\tilde{a}_2$ is dual to the gauge transformation of $a_1$:
\begin{equation}
    \tilde{a}_2 \mapsto \tilde{a}_2 + \delta_{-c-\delta\epsilon}\tilde{\alpha}_1
\end{equation}
while the gauge transformation on $\tilde{c}_2$ is given by:
\begin{equation}
    \tilde{c}_2 \mapsto \tilde{c}_2 + \delta \tilde{\epsilon}_1
\end{equation}
 Using a twisted Leibniz rule constructed in \cite{Benini:2018reh}, it is easy to verify that an off-shell deformation maps the action to:
 \begin{equation}
     I_{\mathbb{D}_k}' = \frac{2\pi}{k}\int_{M_4} \Tilde{a}_2\cup_{c+\delta\epsilon}\delta_{c+\delta\epsilon} a_1 + \pi\int_{M_4} \Tilde{c}_2\cup \delta c_1
 \end{equation}

Finally, it is worth mentioning that such a hierarchy of gauge transformation also appears in higher group gauge theories, which is a more general class of topological sigma models. In fact, the (3+1)D untwisted $\mathbb{D}_k$-DW theory examples in this work are dualized 2-group gauge theories. We refer the readers to \cite{Freed2022FOURLO,Freed:2022qnc,Kapustin:2013uxa} for further details.

\subsection{$\mathbb{D}_4$ Lagrangian} \label{D4-action}
In this section, we present various equivalent $\bbD_4$ Lagrangians. It is known that $\mathbb{D}_4 = (\bbZ_2\times\bbZ_2)\rtimes \bbZ_2$ admits two different split extensions. By the construction introduced in Sec. \ref{charge-conjugation-gauged-action}, the action reads
\begin{equation}
    I_{\bbD_4} = \pi \int \tilde{a}_2 \cup_c \delta_c a_1 + \pi \int \tilde{b}_2 \cup_c \delta_c b_1 + \pi \int \tilde{c}_2 \cup \delta c_1,
    \label{Eq: I_D4}
\end{equation}
where $a_1$, $b_1$ are $\bbZ_2$-valued 1-cochains describing the two components of the Klein-four subgroup $V = \bbZ_2 \times \bbZ_2$, while $c_1$ is the $\bbZ_2$-valued 1-cochain for the quotient sector in the split extension 
\begin{align}
    1\rightarrow V \rightarrow \bbD_4 \rightarrow \bbZ_2\rightarrow 1.
    \label{Eq: split extension}
\end{align}
    Let $V = \mathbb{Z}_2 \times \mathbb{Z}_2$. There are two split extensions with kernel $V$, where the twists in the additive notation act on $V$ as:
\begin{equation}
    \rho_1:(a,b)\mapsto(b,a),\quad \rho_2:(a,b)\mapsto (a,a+b).
\end{equation}
$\rho_1$ exchanges the two $\bbZ_2$ factors, while $\rho_2$ acts on $V$ by a shift. 

These two choices enter the theory only through the twisted coboundary $\delta_c$ and the associated gauge transformations. For simplicity here we only state the off-shell gauge transformation induced by a shift $c_1 \rightarrow c_1 + \delta\epsilon_0$. For the twist $\rho_1$, the gauge fields have the following off-shell deformations:
\begin{equation}
    \begin{split}
        a_1 &\mapsto a_1 + \delta_{c+\delta\epsilon}\alpha_0\\
        b_1 &\mapsto b_1 + 
        \delta_{c+\delta\epsilon}\beta_0\\
        \tilde{a}_2 &\mapsto \tilde{a}_2 + \delta_{-c-\delta\epsilon}\tilde{\alpha}_1\\
        \tilde{b}_2 &\mapsto \tilde{b}_2 + \delta_{-c-\delta\epsilon}\tilde{\beta}_1
    \end{split}
\end{equation}
For the twist $\rho_2$, we have the following off-shell gauge transformations:
\begin{equation}
    \begin{split}
        a_1 &\mapsto a_1 + \delta_{c+\delta\epsilon}\alpha_0\\
        b_1 &\mapsto b_1 + \delta_{}\beta_0\\
        \tilde{a}_2 &\mapsto \tilde{a}_2 + \delta_{}\tilde{\alpha}_1\\
        \tilde{b}_2 &\mapsto \tilde{b}_2 + \delta_{-c-\delta\epsilon}\tilde{\beta}_1
    \end{split}
\end{equation}

    Finally, we note that the Lagrangian of the untwisted $\mathbb{D}_4$-DW theory can also be constructed by gauging a $\mathbb{Z}_2^2$ 0-form symmetry in an untwisted $\mathbb{Z}_2$-DW theory, where the symmetry does not permute the operators but carries nontrivial fractionalization data. The relevant action is a central extension:
    \begin{equation}\label{eq-central-extension}
    0 \rightarrow \mathbb{Z}_2 \rightarrow \mathbb{D}_4 \rightarrow \mathbb{Z}_2^2 \rightarrow 0
\end{equation}
    The $\mathbb{Z}_2^2$ gauged action reads:
    \begin{equation}
    I_{\bbD_4} = \pi \int \tilde{a}_2 \cup \delta a_1 + \pi \int \tilde{b}_2 \cup \delta b_1 + \pi \int \tilde{c}_2 \cup \delta c_1 + \pi\int\tilde{b}_2\cup a_1 \cup c_1
    \label{Eq: I_D4}
\end{equation}
    where $b_1$ is the gauge field of the original untwisted $\mathbb{Z}_2$-DW theory. Integrating out the Lagrange multipliers, the action becomes:
    \begin{equation}\label{eq-central-extension-DW}
        I_{\mathbb{D}_4} = \pi\int \tilde{b}_2\cup a_1\cup c_1
    \end{equation}
    where:
    \begin{equation}
        a_1\cup c_1 \in H^2(\mathbb{Z}_2^2,\mathbb{Z}_2)
    \end{equation}
    is the pullback of the extension class of Eq. \eqref{eq-central-extension} to spacetime. The Lagrangian itself defines a topological sigma model from $M_4$ to the classifying space $\mathbb{B}\mathbb{Z}_2^3$ with a nontrivial topological action. Similar to the previous split extensions, the central extension Eq. \eqref{eq-central-extension} also defines a fibration of classifying spaces:
    \begin{equation}
        B\mathbb{Z}_2^2 \rightarrow B\mathbb{Z}_2^3 \rightarrow B\mathbb{Z}_2
    \end{equation}
    However, it can be shown that the on-shell and off-shell gauge transformations of this theory in the BF formalism agree with each other. One can recover the hierarchy of gauge transformations by appropriately suppressing the small gauge transformation parameters. See \cite{Xue:2025enx} for further details. 
    
\section{Anomaly-Free Conditions} \label{section-anomalies}

In this section, we will work out the obstruction-free conditions for $\mathbb{Z}_2^C$ gauging of $\mathbb{Z}_k$ DW theories in $(3+1)D$ without symmetry fractionalization by anomaly inflow. Especially, we show that all the symmetries we considered in the previous section are anomaly-free. 

As mentioned in Sec. \ref{BF-theory-formulation}, it is convenient to integrate out the Lagrange multipliers for 't Hooft anomaly analysis. We can construct a new $\mathbb{D}_k$ action $\omega_{\mathbb{D}_k}$ in terms of the degrees of freedom of the original $\mathbb{Z}_k$ DW theory as well as the $\mathbb{Z}_2$ background gauge field. The $\mathbb{Z}_2^C$ symmetry is anomaly free only when the background gauge field variation of $\omega_{\mathbb{D}_k}$ can be canceled by $(3+1)$D local counterterms. Because of the ambiguities in $(3+1)$D counterterms, it is more convenient to extend $\omega_{\mathbb{D}_{k}}$ to a 5D manifold $M_5$ with boundary $M_4$ and solve for certain conditions that allow the 4D action to be extended to a trivial 5-cocycle $\omega_5 = \delta \omega_{\mathbb{D}_{k}}$. In particular, these conditions constrain the coupling of background gauge fields with the existing dynamical gauge fields. Because of the anomaly inflow assumption, this critical condition is equivalent to the anomaly-free condition of the  $\mathbb{Z}_2^C$ symmetry of the $(3+1)$D $\mathbb{Z}_k$ DW theory. This framework was first introduced in \cite{Kapustin:2014zva} and it is applicable to all discrete symmetry gauging of DW theories with Lagrangian formulations. If the 0-form symmetry action is organized by a split extension:
\begin{equation}
    0 \rightarrow \mathbb{Z}_k \xrightarrow{\iota} \mathbb{Z}_k\rtimes\mathbb{Z}_2^{C} \rightarrow \mathbb{Z}_2^{C} \rightarrow 0
\end{equation}
then the anomaly free condition admits a much more elegant criterion \cite{Kapustin:2014zva,Muller:2018doa}. The $\mathbb{Z}_2^C$ symmetry is anomaly free when there exists $\omega_{\mathbb{D}_{k}}\in H^4(B\mathbb{D}_{k}, U(1))$ so that $\iota^*\omega_{\mathbb{D}_k} = \omega_{\mathbb{Z}_k}$. 

First we show that the charge-conjugation symmetry action that we considered above is anomaly free. After integrating out the Lagrange multiplier field, we see that the on-shell action of the original $\mathbb{Z}_k$ DW theory and the proposed $\mathbb{Z}_2^C$ gauged theory are both topologically trivial when pulled back to spacetime:
\begin{equation}
    \rho^*\omega_{\mathbb{D}_k} = [0] 
\end{equation}
Therefore, Eq. \eqref{eq-CC-gauged-Lagrangian} is a good BF-Lagrangian for the untwisted $\mathbb{D}_k$ DW theory.

This immediately satisfies the pull-back definition of the anomaly-free condition, because the map $\iota$ in the group extension naturally defines a map of the classifying space $B\mathbb{Z}_k$ and $B{\mathbb{D}_k}$, hence a homomorphism of the cohomology groups $\iota^*: H^{4}(B\mathbb{D}_k,U(1))\rightarrow H^4(B\mathbb{Z}_k,U(1))$. The two trivial topological actions are simply the identity elements of $H^{4}(B\mathbb{D}_k,U(1))$ and $H^4(B\mathbb{Z}_k,U(1))$, which satisfies $\iota^*[0]_{\mathbb{D}_k} = [0]_{\mathbb{Z}_k}$ by definition. We note that this observation can be directly generalized to an $H^{(0)}$ symmetry acting on an untwisted DW theory in arbitrary spacetime dimension with a finite abelian gauge group $A$ via the split extension:
\begin{equation}
    0 \rightarrow A \rightarrow A\rtimes H \rightarrow H \rightarrow 0
\end{equation}
The $H^{(0)}$ symmetry is anomaly free and the $H^{(0)}$ gauged theory is an untwisted $ A\rtimes H$ DW theory. 

Now we consider the symmetry action by central extension. In this case, the on-shell action is given by Eq. \eqref{eq-central-extension-DW}.
Integrating out $b_1$ implements the constraint $\delta \tilde{b}_2 = 0$. Moreover, integrating out $\tilde{a}_2$ and $\tilde{c}_2$ implements the constraints $\delta a_1 = 0$ and $\delta c_1=0$, respectively. Therefore, the extension of the 4-cocycle $\pi\, \tilde b_2\cup a_1 \cup c_1$ to 5D:
\begin{equation}
    \pi\,\delta (\tilde{b}_2 \cup a_1 \cup c_1)
\end{equation}
is necessarily cohomologically trivial by Leibniz rule. Therefore, this symmetry is anomaly-free.

\section{Linking and Higher Gauging Condensation Defect Dressing}\label{section-linkings}

In this section, we will work out the operator formalism of untwisted $\mathbb{D}_{k}$-DW theory. Given an action, one can derive a collection of gauge transformations in the $G$-DW theory and define gauge invariant objects on closed oriented submanifolds of spacetime. We will see that the requirement of gauge invariance of the operators naturally leads to the emergence of higher gauging condensation defects \cite{Roumpedakis:2022aik, Cordova:2024jlk, Cordova:2024mqg}.

The idea of higher gauging condensation defects is to start from a parent gauge theory with an ordinary or higher-form symmetry and gauge the symmetry only along a submanifold rather than in the entire spacetime \cite{Roumpedakis:2022aik}. In finite-group gauge theories, this provides a concrete way to realize both invertible and non-invertible symmetry defects: one first writes a bare operator and then dresses it by suitable condensation factors so that it becomes gauge invariant and has the correct fusion and linking properties. In our construction below, we will repeatedly use condensation defects $\mathcal{S}_c$ and $\tilde{\mathcal{S}}_c$ by higher gauging a $\mathbb{Z}_2$ 1-form symmetry.

Similar to the constructions in our previous work \cite{Xue:2025enx}, the main higher gauging condensation defects involved are the diagonal electric condensation defects $\mathcal{D}_K$ introduced in \cite{Cordova:2024jlk, Cordova:2024mqg}, where $K\triangleleft G$ is a normal subgroup of the spacetime gauge group.  These diagonal condensation defects are orientation reversal invariant with a trivial world-volume topological action, and we refer the readers to \cite{Cordova:2024jlk, Cordova:2024mqg} for their physical implications. In our work, we only need their self-fusion rules
\begin{equation}
    \mathcal{D}_K \times \mathcal{D}_{K} = \frac{\abs{G}}{\abs{K}}\mathcal{D}_{K} 
\end{equation}
We will see that the main role of the electric condensation is to ensure gauge invariance.

A basic diagnostic of the operator spectrum in DW theory is the Hopf link between a Wilson line $W_\rho(S^1)$, labeled by an irreducible representation $\rho$, and a 't Hooft surface $T_{[g]}(S^2)$, labeled by a conjugacy class $[g]$. The linking invariants contain representation-theoretic data of the gauge group. More precisely, the Hopf link equals \cite{Heidenreich:2021xpr}:
\begin{equation}\label{eq-Hopf-Link-eqn}
    \expval{W_{\rho}(S^1)T_{[g]}(S^2)} = \chi_\rho([g]) \times \mathrm{size}([g]). 
\end{equation}
where an overall normalization factor of the DW theory partition function is suppressed. This is the basic relation that allows one to reconstruct the character table from the topological linking data. In the present higher-gauging-condensation-defect construction, the same relation emerges after dressing the bare electric and magnetic operators by the appropriate electric and magnetic condensation defects. 

The physical reason is also straightforward. A Wilson line inserts an electric probe charge transforming in the irrep $\rho$, while a linked 't Hooft surface inserts a magnetic flux in the conjugacy class $[g]$. When the two operators are linked, the electric probe encircles the magnetic flux and picks up the corresponding Aharonov–Bohm phase, or more generally the action of the holonomy $g$ in $\rho$. Taking the trace over the representation gives the character $\chi_\rho(g)$. Since a ’t Hooft operator associated with a non-central conjugacy class is represented by a sum over all elements in the class, the final answer is weighted by the size of conjugacy class.

In fact, generic correlation functions of DW theories are evaluated similarly by unlinking all the operator insertions and then shrinking all the operators to a point. The Aharonov-Bohm phases can be understood as an obstruction to the unlinking process. Particularly, the one-point function of any operator on a sphere measures its quantum dimension up to an overall normalization factor of the partition function. See \cite{Xue:2025enx} for an extended discussion.

In this section, we show how to extract the representation theory information of the gauge group $\mathbb{D}_{k}$ from the linking invariant calculations. Recall that $\mathbb{D}_{k}$ can be represented by the following relation:
\begin{equation}
\mathbb{D}_{k}=\langle r,s \mid r^k=1,\ s^2=1,\ srs^{-1}=r^{-1}\rangle,
\label{Eq: defition of D_k}
\end{equation}
The representations for $k$-even and $k$-odd are slightly different, and we will discuss them separately in Sec. \ref{Subsec: Dk_odd} and Sec. \ref{Subsec: Dk_even}. The case $k=4$ has more equivalent operator formalisms, which we present in Sec. \ref{Subsec: D4}.

\subsection{$\mathbb{D}_{k}$ for $k$-odd \label{Subsec: Dk_odd}}

Recall that $\mathbb{D}_k$ for odd $k$ has the following conjugacy classes:
\begin{equation}
    \{1\},\, \{r^t,r^{-t}\,|\,1\leq t\leq (k-1)/2\},\, \{r^ps\,|\, 0\leq p \leq k-1 \}
\end{equation}
$\mathbb{D}_k$ has two 1D linear irreps $1$ and $\chi_-$, and $(k-1)/2$ 2D linear irreps $\chi_j$, labeled by $j=1,\dots, (k-1)/2$. Similar to usual BF theory formulations, the operator spectrum is constructed by exponentiating the gauge fields. However, subtlety arises when twisted cohomologies are involved.

\begin{figure}[h]
\centering
\begin{tikzpicture}[scale=1]
    \coordinate (i) at (90:1);
    \coordinate (j) at (210:1);
    \coordinate (k) at (330:1);

    \draw[line width=1.2pt]
        (i) -- node[left] {$c_{ij}$} (j)
            -- node[below] {$c_{jk}$} (k)
            -- node[right] {$c_{ki}$} (i);

    \node[above] at (i) {$i$};
    \node[below left] at (j) {$j$};
    \node[below right] at (k) {$k$};

    \draw[->, line width=1.2pt] ($(i)!0.35!(j)$) -- ($(i)!0.65!(j)$);
    \draw[->, line width=1.2pt] ($(j)!0.35!(k)$) -- ($(j)!0.65!(k)$);
    \draw[->, line width=1.2pt] ($(k)!0.35!(i)$) -- ($(k)!0.65!(i)$);
\end{tikzpicture}
\caption{A triangulation of \(S^1\).}
\label{fig-circle-triangulation}
\end{figure}
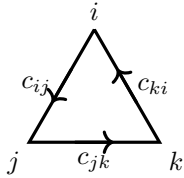

Let us first consider the line operators.   Gauging a $\mathbb{Z}_2^C$ symmetry without fractionalization always leads to a nontrivial $\mathbb{Z}_2$ Wilson line:
\begin{equation}
    U_c(M_1) = e^{i\oint_{M_1}c_1}
\end{equation}
where the flux integral is an abbreviation of the evaluation of a 1-cocycle $c_1$ on a 1-cycle $M_1$. For example, when $M_1$ is $S^1$ as in Fig. \ref{fig-circle-triangulation}, the integral represents:
\begin{equation}
    \oint_{M_1}c_1 = c_1([M_1]) = c_{ij} + c_{jk} + c_{ki}
\end{equation}

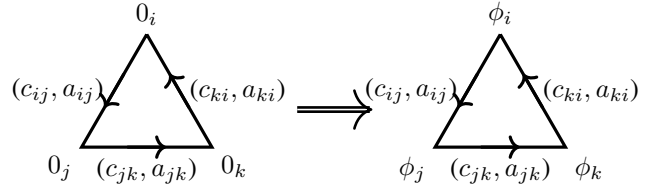
\begin{figure}[h]
\centering
\begin{tikzpicture}[scale=1]

    \begin{scope}[xshift=0cm]
        \coordinate (iL) at (90:1);
        \coordinate (jL) at (210:1);
        \coordinate (kL) at (330:1);

        \draw[line width=1.2pt]
            (iL) -- node[left] {($c_{ij}, a_{ij}$)} (jL)
                 -- node[below] {($c_{jk}, a_{jk}$)} (kL)
                 -- node[right] {($c_{ki}, a_{ki}$)} (iL);

        \node[above] at (iL) {$0_i$};
        \node[below left] at (jL) {$0_j$};
        \node[below right] at (kL) {$0_k$};

        \draw[->, line width=1.2pt] ($(iL)!0.35!(jL)$) -- ($(iL)!0.65!(jL)$);
        \draw[->, line width=1.2pt] ($(jL)!0.35!(kL)$) -- ($(jL)!0.65!(kL)$);
        \draw[->, line width=1.2pt] ($(kL)!0.35!(iL)$) -- ($(kL)!0.65!(iL)$);
    \end{scope}

    \begin{scope}[xshift=4.7cm]
        \coordinate (iR) at (90:1);
        \coordinate (jR) at (210:1);
        \coordinate (kR) at (330:1);

        \draw[line width=1.2pt]
            (iR) -- node[left] {($c_{ij}, a_{ij}$)} (jR)
                 -- node[below] {($c_{jk}, a_{jk}$)} (kR)
                 -- node[right] {($c_{ki}, a_{ki}$)} (iR);

        \node[above] at (iR) {$\phi_i$};
        \node[below left] at (jR) {$\phi_j$};
        \node[below right] at (kR) {$\phi_k$};


        \draw[->, line width=1.2pt] ($(iR)!0.35!(jR)$) -- ($(iR)!0.65!(jR)$);
        \draw[->, line width=1.2pt] ($(jR)!0.35!(kR)$) -- ($(jR)!0.65!(kR)$);
        \draw[->, line width=1.2pt] ($(kR)!0.35!(iR)$) -- ($(kR)!0.65!(iR)$);
    \end{scope}

    \draw[->, double, line width=1pt]
        (2.0,0) -- node[above] {} (3.0,0);

\end{tikzpicture}
\caption{An on-shell gauge transformation of $e^{i\oint_{S^1}a_1}$ on a loop with nontrivial $c_1$ profile. $0_i, 0_j, 0_k$ means that there are no data associated to the sites $i,j,k$. }
\label{fig-gauge-transformation}
\end{figure}

One may attempt to define the object $e^{i\oint_{M_1}a_1}$, but it is not gauge invariant under the on-shell gauge transformation Eq. \eqref{eq-on-shell deformation}. To see this, let $M_1=S^1$. With the triangulation in Fig. \ref{fig-circle-triangulation}, an on-shell gauge transformation over an $S^1$ with a nontrivial $c_1$ background shifts the integral by:
\begin{equation}
        \oint_{S^1}\delta_c\alpha_0= \rho(c_{ij})\alpha_j - \alpha_i + \rho(c_{jk})\alpha_k - \alpha_j + \rho(c_{ki})\alpha_i-\alpha_k
\end{equation}
This means that the sites $i,j,k$ now host local gauge transformation parameters:
\begin{equation}
    \begin{split}
    \phi_i  &=\rho(c_{ki})\alpha_i - \alpha_i\\
    \phi_j  &=\rho(c_{ij})\alpha_j - \alpha_j\\
        \phi_k  &=\rho(c_{jk})\alpha_k - \alpha_k
    \end{split}
\end{equation}
See Fig. \ref{fig-gauge-transformation} for a demonstration.
Gauge invariance requires the cancellation of these parameters, and this can only be achieved by trivializing $c_1$ over the $S^1$. The same reasoning applies to off-shell gauge transformations, where on a link $ij$ replace $c_{ij}$ with:
\begin{equation}
    c_{ij}' = (c_1+\delta\epsilon_0)_{ij} = c_{ij} + \epsilon_j-\epsilon_i
\end{equation}
Therefore, all we need is the trivialization of cohomology classes of $c_1$.

A simple way to trivialize $[c_1]$ on any loop is by stacking with the following delta function:
\begin{equation}
    \delta_c(M_1) = \frac{1}{2}(1 + e^{i\oint_{M_1} c_1})
\end{equation}
Moreover, since the $\mathbb{Z}_2^C$ symmetry exchanges the gauge field $a_1$ with $-a_1$, the true physical object should be the following orbit of the $\mathbb{Z}_2^C$ action:
\begin{equation}
    \hat{U}_a^{(1)}= \left(e^{i\oint a_1} + e^{-i\oint a_1}\right)\delta_c
\end{equation}
In fact, in the language of \cite{Cordova:2024jlk, Cordova:2024mqg}, one should formally rewrite the operator as:
\begin{equation}
    \hat{U}_a^{(1)}(M_1)= \frac{e^{i\oint a_1} + e^{-i\oint a_1}}{2} \mathcal{S}_c(M_1)
\end{equation}
where 
\begin{equation}
    \mathcal{S}_c(M_1) = 2\delta_c(M_1) 
    \label{eq: definition_Sc}
\end{equation}
is the electric condensation defect associated with the $\mathbb{Z}_k$ normal subgroup of the spacetime gauge group $\mathbb{D}_k$. In this notation, the higher gauging condensation defect $\mathcal{S}_c$ carries the quantum dimension of the operator:
\begin{equation}
    \expval{\hat{U}_a^{(1)}} = 2
\end{equation}
$\mathbb{D}_k$ has $\frac{k-1}{2}$ such linear irrep in total, and they are similarly labeled by:
\begin{equation}
    \hat{U}_a^{(j)}= \frac{e^{i\oint j a_1} + e^{-i\oint j a_1}}{2}\mathcal{S}_{c}
\end{equation}
where $j\in \{1,2,\dots,\frac{k-1}{2}\}$. Together with the trivial irrep, we have the $1+1+\frac{k-1}{2} = \frac{k+3}{2}$ irreps of $\mathbb{D}_{k}$ in total. 

Now we consider the 't Hooft operators. The only invertible surface operator is the trivial surface. We can attempt to define an object $e^{i\oint_{M_2}\tilde{a}_2}$, but the same gauge invariance consideration requires us to trivialize all cohomology classes of $[c_1]$ on all possible 1-cycles of $M_2$, which is achieved by the stacking of the following operator:
\begin{equation}
    \Delta_c(M_2) = \prod_{\gamma\in H_1(M_2,\mathbb{Z}_2)}\delta_c(\gamma) = \frac{1}{\abs{H_1(M_2,\mathbb{Z}_2)}}\sum_{\gamma\in H_1(M_2,\mathbb{Z}_2)} U_c(\gamma) 
\end{equation}

Now we can construct surface operators with quantum dimension 2, which can be understood as the size-2 conjugacy classes of the $\mathbb{Z}_2^C$ orbits of $\mathbb{Z}_k \triangleleft \mathbb{D}_{k}$
\begin{equation}
    \hat{U}_{\Tilde{a}}^{(t)}= \frac{e^{it\oint \Tilde{a}_2} + e^{-it\oint \Tilde{a}_2}}{2} \tilde{\mathcal{S}}_c = \left(e^{it\oint \Tilde{a}_2} + e^{-it\oint \Tilde{a}_2}\right)\Delta_c
\end{equation}
where $t\in\{1,2,\dots,\frac{k-1}{2}\}$ and $\mathcal{S}_c = 2\Delta_c$ is a codimension-2 electric condensation defect associated to $\mathbb{Z}_k\triangleleft \mathbb{D}_k$. Finally, consider the object $e^{i\oint_{M_2}\tilde{c}_2}$, which should correspond to the remaining size $k$ conjugacy class. The quantum dimension can be provided by the operator:
\begin{equation}
  M_{\tilde a}(M_2) = \sum_{l=0}^{k-1} e^{il\oint_{M_2} \Tilde{a}_2}
\end{equation}
which is not gauge invariant on its own. Therefore, we need a further stacking of $\mathcal{S}_c$, which leads us to the following representation of the conjugacy class $[s]$: 
\begin{equation}
    \hat{U}_{\Tilde{c}} = \frac{1}{2}e^{i\oint \Tilde{c}} \tilde{\S}_c M_{\tilde{a}},  
\end{equation}
Similar to the construction of non-invertible Wilson lines, the operator $e^{i\oint_{M_2}\tilde{c}_2}M_{\tilde{a}}$ can
be understood as a size $k$-orbit of $\mathbb{D}_k$ under conjugation action, which needs to be dressed with $\tilde{S}_c$ to ensure gauge invariance. Together, this gives $1 + \frac{k-1}{2}+1 = \frac{k+3}{2}$ conjugacy classes as expected. 

We end our discussion of untwisted $\mathbb{D}_{k}$ by explicitly carrying out a few linking invariant calculations.  Let's consider the Hopf link between $\hat{U}_a^{(j)}(S^1)$ and $\hat{U}_{\tilde a}^{(t)}(S^2)$. Since 
$a_1$ serves as a source of $\tilde{a}_2$, integrating out $a_1$ in the path integral converts the $\tilde a_2$ monodromy into Aharonov-Bohm phases. Meanwhile, since nothing sources the $c_1$ fields in the electric condensation $\mathcal{S}_c$, we are free to shrink $\mathcal{S}_c$ to a point. Altogether:
\begin{align}
    &\expval{\hat{U}_a^{(j)}(S^1)\hat{U}_{\tilde a}^{(t)}(S^2)} \nonumber \\
    =& \expval{ e^{i\oint j a_1}(e^{it\oint \Tilde{a}_2} + e^{-it\oint \Tilde{a}_2})\frac{\mathcal{S}_c}{2}} + \mathrm{c.c.} \nonumber \\
    =& (e^{\frac{2\pi i}{k}jt} + e^{-\frac{2\pi i}{k}jt}) + \mathrm{c.c.} \nonumber \\
    = & 4 \cos\Big(\frac{2\pi i}{k}jt\Big) \nonumber \\
    = & \chi_j([r^t])\times\text{size}([r^t])
\end{align}

Similarly, consider the Hopf link between $\hat{U}_a^{(j)}(S^1)$ and $\hat{U}_{\tilde{c}}(S^2)$. In this case, the $\tilde c_2$ field on $\hat{U}_{\tilde{c}}(S^2)$ nontrivially sources the $c_1$ fields, whose monodromies are condensed on $S^1$ by the electric condensation $\mathcal{S}_c$. Therefore, we have
\begin{align}
    &\expval{\hat{U}_a^{(j)}(S^1) \hat{U}_{\tilde{c}}(S^2)} \nonumber \\
    =& \frac{1}{2} \expval{(e^{i\oint_{S^1} j a_1} + e^{-i\oint_{S^1} j a_1})(1+U_c(S^1))e^{i\oint_{S^2}\tilde{c}_2} \mathcal{S}_c(S^2)} \nonumber  \\
    =&  \expval{(e^{i\oint_{S^1} j a_1} + e^{-i\oint_{S^1} j a_1})(1+e^{\frac{2\pi i}{2}}) } \nonumber \\
    =& 0
\end{align}
as expected, where we shrunk $\tilde{\mathcal{S}}_c(S^2)$ in the third line. The remaining entries of the character table can be reconstructed similarly and the results are recorded in Appendix \ref{appendix-linking-results}.

\subsection{$\mathbb{D}_{k}$ for $k$-even \label{Subsec: Dk_even}}

The $k$-even case contains more families of operators than the $k$-odd case, but the idea is the same. Let us set $k=2m$. The conjugacy classes are:
\begin{equation}
    \{1\},\, \{r^m\}, \, \{r^t,r^{-t}\}, \, \{r^ps\,|\,p\text{-even}\}, \, \{r^ps\,|\,p\text{-odd}\},
\end{equation}
where $1\leq t\leq m-1$ and the last two are both size $m$ conjugacy classes. $\mathbb{D}_{2m}$ has four 1D irreps 1, $\chi_{+-}, \chi_{-+}, \chi_{--}$ and $m-1$ 2D irreps labeled by $\chi_j$, where $1\leq j \leq m-1$. 

Let us first work out the Wilson lines. First we have the trivial line $\mathbf{1}$ and $U_c = e^{i\oint c}$, which can be identified with $\mathbf 1$ and $\chi_{+-}$. We also have the gauge invariant object:
\begin{equation}
    U_a(M_1) = e^{i\oint_{M_1} ma_1}
\end{equation}
This is because a gauge transformation:
\begin{equation}
     ma_1 \mapsto ma_1 + \delta_c (m\alpha_0)
\end{equation}on $ma_1$ shifts it by an ordinary coboundary. In terms of the component form:
\begin{equation}
    \begin{split}
       \Big(\delta_c (m\alpha_0)\Big)_{ij} =& \rho(c_{ij})(m\alpha_0)_j - (m\alpha_0)_i\\
       =&(m\alpha_0)_j- (m\alpha_0)_i\\
       =& \delta(m\alpha_0)_{ij}
    \end{split}
\end{equation}
where we have used the fact that the charge conjugation action represented by the twist $\rho(c_{ij})$ acts trivially on $m\in \mathbb{Z}_{k}$. Therefore, this operator does not need an $\mathcal{S}_c$ dressing. Fusing $U_a$ with $U_c$ creates another invertible Wilson line:
\begin{equation}
    U_{a,c} = U_a\times U_c
\end{equation}
which we identify as the irrep $\chi_{--}$.
The non-invertible line construction is analogous to the $k$-odd case, so we directly state the result:
\begin{equation}
    \hat{U}_a^{(j)} = \frac{e^{i\oint ja_1} + e^{-i\oint ja_1}}{2} \mathcal{S}_c (M_1)
\end{equation}
where $1\leq j\leq m-1$. 

The 't Hooft surface construction is analogous to the $k$-odd case. The only difference is the dressing of the magnetic condensations responsible for the quantum dimensions of the operators. There are only two invertible surfaces:
\begin{equation}
    \mathds{1} \quad \text{and}\quad U_{\Tilde{a}} = e^{im\oint \Tilde{a}_2}
\end{equation}
There are $m-1$ non-invertible surfaces of quantum dimension 2, corresponding to the $m-1$ size-2 conjugacy classes:
\begin{equation}
    \hat{U}_{\Tilde{a}}^{(r)} = \frac{e^{i\oint r\Tilde{a}} + e^{-i\oint r\Tilde{a}}}{2} \tilde{\mathcal{S}}_c
\end{equation}
where $r\in \{1,2,\dots, m-1\}$. The remaining two size $m$ conjugacy classes are given by constructing gauge invariant objects with $e^{i\oint_{M_2}\tilde{c}_2}$. There are two natural non-gauge-invariant objects with quantum dimension $m$:
\begin{equation}
    \begin{split}
            M_{[s]} &= \sum_{l=0}^{m-1}e^{2i\oint l\tilde{a}_2} \\
            M_{[rs]} &= \sum_{l=0}^{m-1}e^{i\oint (2l+1)\tilde{a}_2}
    \end{split}
\end{equation}
They can be made gauge invariant by further dressing with the higher gauging condensation defect $\Delta_c = \frac{1}{2}\tilde{S}_c$, which is an electric condensation defect associated to the $\mathbb{Z}_k$ normal subgroup of $\mathbb{D}_{k}$. Therefore, the correct operators are:
\begin{equation}
    \begin{split}
        \hat{U}_{\tilde{c}}^{[s]} &= \frac{1}{2}e^{i\oint_{M_2}\tilde{c}_2}\tilde{\mathcal{S}}_c M_{[s]}\\
        \hat{U}_{\tilde{c}}^{[rs]} &= \frac{1}{2}e^{i\oint_{M_2}\tilde{c}_2}\tilde{\mathcal{S}}_c M_{[rs]}
    \end{split}
\end{equation}

\subsection{Equivalent Formulations of Untwisted $\mathbb{D}_{4}$ Gauge Theories}\label{Subsec: D4}

Given the $\bbD_4$-DW theory (Eq. \eqref{Eq: I_D4}) in Sec. \ref{D4-action}, in this section, we construct the corresponding Wilson and 't Hooft operators using the higher gauging condensation defects. The bare $\bbZ_2\times\bbZ_2$ variables and the dressing factors will depend on which split extension presentation we use as introduced in Sec. \ref{D4-action}. For the operator content of the central extension, see \cite{Xue:2025enx,Putrov:2016qdo,He:2016xpi, Robbins:2025puq, Bergman:2024its, Bergman:2026lnz} for further details.

We first consider the representative $\rho_1$, where $\bbD_4$ is defined as
\begin{equation}
    \bbD_4=\braket{a,b,t\,|\, a^2=b^2=t^2=1,ab=ba,tat=b,tbt=a}.
    \label{Eq: definition_split_D4}
\end{equation}
The conjugacy classes are given as
\begin{equation}
    \{1\}, \, \{ab\}, \, \{a,b\}, \, \{t, abt\}, \, \{at, bt\}.
\end{equation}
$\bbD_4$ has four 1D irreps: 1, $\chi_{+-}$, $\chi_{-+}$, $\chi_{--}$, and one 2D irrep

Following the same procedure in Sec. \ref{Subsec: Dk_odd} and \ref{Subsec: Dk_even}, the four invertible Wilson lines are 
\begin{align}
    & \mathds{1},\quad U_c(M_1) = e^{i\oint c_1},\quad U_{a,b}(M_1) = e^{i\oint a_1+b_1}, \nonumber \\
    & U_{a,b,c}(M_1) = e^{i\oint a_1+b_1+c_1},
\end{align}
and one non-invertible line is
\begin{equation}
    \hat{U}_{a} = \frac{1}{2}\left(e^{i\oint a_1} + e^{i\oint b_1} \right)\S_c,
\end{equation}
where $\S_c$ is defined by Eq. \eqref{eq: definition_Sc}.

Similarly, the 't Hooft surfaces can be constructed as follows. The two invertible surfaces correspond to the size-1 conjugacy classes:
\begin{equation}
    \mathds{1}, \quad U_{\tilde{a},\tilde{b}} = e^{i\oint \tilde{a}_2 + \tilde{b}_2}.
\end{equation}
The remaining surfaces are non-invertible
\begin{align}
    \hat{U}_{\tilde{a}} = & \frac{1}{2} e^{i\oint \tilde{a}_2}  \tilde{\S}_c M_{[ab]}, \\
    \hat{U}_{\tilde{c}} = & \frac{1}{2}e^{i\oint \tilde{c}_2} \tilde{\S}_c M_{[ab]}, \\
    \hat{U}_{\tilde{a}, \tilde{c}} = & \frac{1}{2} e^{i\oint \tilde{a}_2 + \tilde{c}_2} \tilde{\S}_c M_{[ab]},
\end{align}
where $M_{[ab]} = (1+e^{i\oint \tilde{a}_2 + \tilde{b}_2})/2$. 

Now we consider another choice of split extension by replacing the twist conditions with $tat^{-1}=a$ and $tbt^{-1}=ab$. Again, there are four 1D irreps labeled by the signs of $b$ and $t$ and one 2D irrep. The conjugacy classes are given as
\begin{equation}
    \{1\},\, \{a\}, \, \{b,ab\}, \, \{t,at\}, \, \{bt, abt\}.
\end{equation}
The four invertible Wilson lines are
\begin{align}
    & \mathbf{1}, \quad U_c(M_1) = e^{i\oint c_1}, \quad U_b(M_1) = e^{i\oint b_1},  \nonumber \\
    & U_{b,c}(M_1) = e^{i\oint b_1+c_1},
\end{align}
and one non-invertible line is
\begin{equation}
    \hat{U}_a = \frac{1}{2}\left(e^{i\oint a_1} + e^{i\oint a_1+b_1}\right)\S_c.
\end{equation}
The 't Hooft surfaces are listed as follows:
\begin{align}
    &\mathds{1},\quad U_{\tilde{a}} = e^{i\oint \tilde{a}_2}, \quad \hat{U}_{\tilde{b}} = \frac{1}{2}e^{i\oint \tilde{b}_2} \tilde{\S}_c M_a, \nonumber \\
    &\hat{U}_{\tilde{c}} = \frac{1}{2}e^{i\oint \tilde{c}_2} \tilde{\S}_c M_a,\quad \hat{U}_{\tilde{a},\tilde{c}} = \frac{1}{2} e^{i\oint \tilde{a}_2 + \tilde{c}_2} \tilde{\S}_c M_a,
\end{align}
where $M_a = (1 + e^{i\oint \tilde{a}_2})/2$.

Following the previous discussions in Sec. \ref{Subsec: Dk_odd} and \ref{Subsec: Dk_even}, we can show that all these operators give the correct quantum dimensions and Hopf links, thus both of them reproduce the same untwisted $\bbD_4$ DW theory.

\section{Conclusion}\label{section-conclusion}

In this work, we outlined a procedure for constructing BF-Lagrangians for a $G$-DW theory by gauging an abelian $H^{(0)}$ symmetry in abelian $A$-DW theory:
\begin{equation}
    1 \rightarrow A \rightarrow G \rightarrow H \rightarrow 1
\end{equation}
We explicitly constructed the operator spectrum with appropriate higher gauging condensation defects for $G = \mathbb{D}_k$ and verified our Lagrangian construction by matching the linking invariant calculation with the character table of $G$. 

As advertised in \cite{Freed:2022qnc,Freed2022FOURLO}, DW theories are a simple class of a broader class of topological sigma models $M_D \rightarrow X$, where $X$ has a finite number of non-vanishing homotopy groups. When the only non-vanishing homotopy group of $X$ is its fundamental group, we recover the ordinary DW theory. For a more general target space $X$ with a finite number of non-vanishing higher homotopy group, the corresponding topological sigma models are typically called higher group gauge theories. If a BF formulation of the higher group gauge theory is available, it should be a rather straightforward exercise to construct gauge invariant objects from gauge transformations of the BF-action. However, a precise operator formalism of the theory would also require the identification of appropriate defects stacked on the extended operators to ensure gauge invariance. Similar to our previous work \cite{Xue:2025enx}, the stacked higher gauging condensation defects contain nontrivial information about the quantum dimensions of the operators. For future directions, it would be interesting to study the BF-type Lagrangians of more general higher group gauge theories and the corresponding operator formalism, which would provide a more physical approach to studying exotic bosonic topological orders as well as symTFT/symTO/topological holography for a wider class of non-invertible/ higher group global symmetries.

\begin{acknowledgements}
We thank Ken Intriligator and John McGreevy for helpful discussions.
    E.Y.Y. is supported by Simons Foundation award 568420 (Simons Investigator) and award
888994 (The Simons Collaboration on Global Categorical Symmetries). 
\end{acknowledgements}

\appendix

\section{Homotopy Theory Basics}\label{appendix-homotopy}

In this appendix, we quickly review some concepts of homotopy theory involved in the main text. Homotopy theory is the study of spaces and structured objects up to continuous deformations. The main topic of concern is maps between topological spaces and oftentimes it is also convenient to present concepts in terms of commutative diagrams.

Now consider a map $p: E\rightarrow B$ and another map $g: X\rightarrow E$. As mentioned in the main text, we can always compose a third map $g_*p = p\circ g$ so that the following diagram commutes. 
\begin{equation}
    \begin{tikzcd}
        & E \arrow[d,"p"] \\
        X \arrow[ur,"g"] \arrow[r,"\tilde g_*p = p\circ g"'] & B
    \end{tikzcd}
\end{equation}
However, given a map $f:X \rightarrow B$ and a map $p: E\rightarrow B$, it is not guaranteed that one can find a third map $\tilde{f}: X\rightarrow E$ so that $f = p\circ\tilde{f}$ and the following diagram commutes:
\begin{equation}
    \begin{tikzcd}
        & E \arrow[d,"p"] \\
        X \arrow[ur, "\exists\,\tilde{f}\,?"] \arrow[r,"f "] & B
    \end{tikzcd}
\end{equation}
If one can find such an $\tilde{f}$, then it is called a \textit{lift} of the map $f: X \rightarrow B$ through $p$. For example, consider the group extension:
\begin{equation}
    0 \rightarrow \mathbb{Z}_2 \xrightarrow{\iota}\mathbb{Z}_4 \xrightarrow{p} \mathbb{Z}_2' \rightarrow 0
\end{equation}
Let $O$ be the trivial group homomorphism sending all elements of $\mathbb{Z}_2$ to the identity element of $\mathbb{Z}_2'$, then $\iota: \mathbb{Z}_2 \rightarrow\mathbb{Z}_4$ would be a lift of the homomorphism $O$ through $p$. On the other hand, consider the identity map $\text{id}:\mathbb{Z}_2 \rightarrow\mathbb{Z}_2'$ and $p:\mathbb{Z}_4 \rightarrow \mathbb{Z}_2'$, and there does not exist any homomorphism lifting $\text{id}$ through $p$.

Let's consider a particular space $X$ and a special map $p: E\rightarrow B$ so that for any map $f:X\rightarrow B$ there exists a lift $\tilde{f}$ of $f$ through $p$. If for any homotopy $f_t: X\times I \rightarrow B$ of $f$, there exists a map $\tilde{f}_t: X\times I \rightarrow E$ so that the following diagram commutes:
\begin{equation}
    \begin{tikzcd}
        & E \arrow[d,"p"] \\
        X\times I \arrow[ur,"\tilde{f}_t"] \arrow[r,"f_t"] & B
    \end{tikzcd}
\end{equation}
and $\tilde{f}_0 = \tilde{f}$, we say that $p$ has the \textit{homotopy lifting property} for $X$. Namely, we relaxed the triangle diagram for lifting by smearing $f$ and $\tilde{f}$ with homotopies up to compatibility conditions. Furthermore, if the map $p:E \rightarrow B$ satisfies the homotopy lifting property for all spaces $X$, then $p: E\rightarrow B$ is called a \textit{fibration}, where $E$ is the total space and $B$ is the base space. Let $b_0\in B$ be a basepoint, then the pre-image of the base point $F = p^{-1}(b_0)$ in $E$ is a \textit{typical fiber}. In this context, the section of a fibration $p:E \rightarrow B$ is simply a map $s$ that partially inverts $p$ in a sense that $p\circ s = \text{id}_B$. Finally, we introduce a method to construct new fibrations from maps onto the base space $B$.  For any map $f: X \rightarrow B$, we can construct a space:
\begin{equation}
    f^*E = X\times_BE = \{(x,e)\in X\times E\,|\, f(x)=p(e) \}
\end{equation}
The map $f^*p: E\rightarrow X$ acts as a projection $f^*p(x,e) = x$ called the \textit{pullback fibration}. Concretely, we have the following commutative square diagram:
\begin{equation}
    \begin{tikzcd}
f^*E \arrow[r] \arrow[d,"f^*p"'] & E \arrow[d,"p"] \\
X \arrow[r,"f"'] & B
\end{tikzcd}
\end{equation}
To recover the familiar concepts of smooth fiber bundles, one needs to add additional constraints on homotopy fibrations. If we further require that $E$ is locally a product space $p^{-1}(U) \simeq U \times F$, where $U\subset B$ and $F$ is the typical fiber, then a homotopy fibration becomes a fiber bundle. Further demanding smooth structures on $E,F,B$ and extra compatibility conditions recovers the definition of smooth fiber bundles.

\section{Linking Invariant Calculations}\label{appendix-linking-results}

In this appendix, we record the Hopf-link calculations in untwisted $\mathbb{D}_k$-DW theories from path integrals summarized in Table. \ref{tab:linking-k-odd} and \ref{tab:linking-k-even}. The character table information can be extracted by matching with Eq. \eqref{eq-Hopf-Link-eqn} 

\begin{table}[htb]
\centering
\caption{Linking invariants for $k$-odd.}
\renewcommand{\arraystretch}{1.15}
\setlength{\tabcolsep}{8pt}
\[
\begin{array}{c|ccc}
\langle W_{\rho}(S^1)\,T_g(S^2)\rangle
& \mathds{1} & U_{\hat a}^{(t)} & \hat U_{\hat c}
\\ \hline
\mathbf{1} & 1 & 2 & 2 \\
\epsilon    & 1 & 2 & -2 \\
\chi_j      & 2 & 4\cos\!\left(\frac{2\pi jt}{k}\right) & 0
\end{array}
\]
\label{tab:linking-k-odd}
\end{table}

\begin{table}[htb]
\centering
\caption{Linking invariants for $k$-even.}
\renewcommand{\arraystretch}{1.15}
\setlength{\tabcolsep}{7pt}
\[
\begin{array}{c|ccccc}
\langle W_{\rho}(S^1)\,T_g(S^2)\rangle
& \mathds{1} & U_{\hat a} & \hat U_{\hat a}^{(r)} & \hat U_{\tilde{c}}^{[s]}  & \hat U_{\tilde{c}}^{[rs]}
\\ \hline
\mathbf{1}   & 1 & 1 & 2 & m & m \\
\chi_{+-}    & 1 & 1 & 2 & -m & -m \\
\chi_{-+}    & 1 & (-1)^m & 2(-1)^r & m & -m \\
\chi_{--}    & 1 & (-1)^m & 2(-1)^r & -m & m \\
\chi_j       & 2 & 2(-1)^j & 4\cos\!\left(\frac{2\pi jr}{k}\right) & 0 & 0
\end{array}
\]
\label{tab:linking-k-even}
\end{table}

\bibliography{refs}

@article{Kapustin:2014zva,
    author = "Kapustin, Anton and Thorngren, Ryan",
    title = "{Anomalies of discrete symmetries in various dimensions and group cohomology}",
    journal = "{}", 
    eprint = "{1404.3230}",
    archivePrefix = "arXiv",
    primaryClass = "hep-th",
    month = "4",
    year = "2014"
}

@article{Dijkgraaf:1989pz,
    author = "Dijkgraaf, Robbert and Witten, Edward",
    title = "{Topological Gauge Theories and Group Cohomology}",
    reportNumber = "THU-89-9, IASSNS-HEP-89-33",
    doi = "10.1007/BF02096988",
    journal = "Commun. Math. Phys.",
    volume = "129",
    pages = "393",
    year = "1990"
}

@article{Baez:1995xq,
    author = "Baez, J. C. and Dolan, J.",
    title = "{Higher dimensional algebra and topological quantum field theory}",
    eprint = "q-alg/9503002",
    archivePrefix = "arXiv",
    doi = "10.1063/1.531236",
    journal = "J. Math. Phys.",
    volume = "36",
    pages = "6073--6105",
    year = "1995"
}

@article{Lurie2009OnTC,
  title={On the Classification of Topological Field Theories},
  author={Jacob Lurie},
  journal={arXiv: Category Theory},
  year={2009},
  url={https://api.semanticscholar.org/CorpusID:115162503}
}

@article{Chen:2011pg,
    author = "Chen, Xie and Gu, Zheng-Cheng and Liu, Zheng-Xin and Wen, Xiao-Gang",
    title = "{Symmetry protected topological orders and the group cohomology of their symmetry group}",
    eprint = "1106.4772",
    archivePrefix = "arXiv",
    primaryClass = "cond-mat.str-el",
    doi = "10.1103/PhysRevB.87.155114",
    journal = "Phys. Rev. B",
    volume = "87",
    number = "15",
    pages = "155114",
    year = "2013"
}

@book{HatcherAT,
  author    = {Hatcher, Allen},
  title     = {Algebraic Topology},
  publisher = {Cambridge University Press},
  address   = {Cambridge},
  year      = {2002},
  isbn      = {9780521795401},
}

@article{Delcamp:2019fdp,
    author = "Delcamp, Clement and Tiwari, Apoorv",
    title = "{On 2-form gauge models of topological phases}",
    eprint = "1901.02249",
    archivePrefix = "arXiv",
    primaryClass = "hep-th",
    doi = "10.1007/JHEP05(2019)064",
    journal = "JHEP",
    volume = "05",
    pages = "064",
    year = "2019"
}

@inproceedings{Freed2022FOURLO,
  title={FOUR LECTURES ON FINITE SYMMETRY IN QFT},
  author={Daniel S. Freed and Constantin Teleman and Gregory Moore and Daniel S. Freed},
  year={2022},
  url={https://api.semanticscholar.org/CorpusID:250520740}
}

@article{Xue:2025enx,
    author = "Xue, Yuan and Yang, Eric Y. and Zhang, Zipei",
    title = "{On Gauging Finite Symmetries by Higher Gauging Condensation Defects}",
    eprint = "2512.22440",
    archivePrefix = "arXiv",
    journal = "",
    primaryClass = "hep-th",
    month = "",
    year = "2025"
}

@article{Kaidi:2022cpf,
    author = "Kaidi, Justin and Ohmori, Kantaro and Zheng, Yunqin",
    title = "{Symmetry TFTs for Non-invertible Defects}",
    eprint = "2209.11062",
    archivePrefix = "arXiv",
    primaryClass = "hep-th",
    doi = "10.1007/s00220-023-04859-7",
    journal = "Commun. Math. Phys.",
    volume = "404",
    number = "2",
    pages = "1021--1124",
    year = "2023"
}

@article{Muller:2018doa,
    author = {M{\"u}ller, Lukas and Szabo, Richard J.},
    title = "{{\textquoteright}t Hooft Anomalies of Discrete Gauge Theories and Non-abelian Group Cohomology}",
    eprint = "1811.05446",
    archivePrefix = "arXiv",
    primaryClass = "hep-th",
    reportNumber = "EMPG-18-23",
    doi = "10.1007/s00220-019-03546-w",
    journal = "Commun. Math. Phys.",
    volume = "375",
    number = "3",
    pages = "1581--1627",
    year = "2019"
}

@article{Cordova:2024jlk,
    author = "Cordova, Clay and Costa, Davi Bastos and Hsin, Po-Shen",
    title = "{Non-invertible symmetries in finite-group gauge theory}",
    eprint = "2407.07964",
    archivePrefix = "arXiv",
    primaryClass = "cond-mat.str-el",
    doi = "10.21468/SciPostPhys.18.1.019",
    journal = "SciPost Phys.",
    volume = "18",
    number = "1",
    pages = "019",
    year = "2025"
}

@article{Cordova:2024mqg,
    author = "Cordova, Clay and Costa, Davi B. and Hsin, Po-Shen",
    title = "{Non-Invertible Symmetries as Condensation Defects in Finite-Group Gauge Theories}",
    journal="",
    eprint = "2412.16681",
    archivePrefix = "arXiv",
    primaryClass = "cond-mat.str-el",
    month = "12",
    year = "2024"
}

@article{Kapustin:2014gua,
    author = "Kapustin, Anton and Seiberg, Nathan",
    title = "{Coupling a QFT to a TQFT and Duality}",
    eprint = "1401.0740",
    archivePrefix = "arXiv",
    primaryClass = "hep-th",
    doi = "10.1007/JHEP04(2014)001",
    journal = "JHEP",
    volume = "04",
    pages = "001",
    year = "2014"
}

@article{Freed:2022qnc,
    author = "Freed, Daniel S. and Moore, Gregory W. and Teleman, Constantin",
    title = "{Topological symmetry in quantum field theory}",
    journal="",
    eprint = "2209.07471",
    archivePrefix = "arXiv",
    primaryClass = "hep-th",
    month = "9",
    year = "2022"
}

@article{Kitaev:1997wr,
    author = "Kitaev, A. Yu.",
    title = "{Fault tolerant quantum computation by anyons}",
    eprint = "quant-ph/9707021",
    archivePrefix = "arXiv",
    doi = "10.1016/S0003-4916(02)00018-0",
    journal = "Annals Phys.",
    volume = "303",
    pages = "2--30",
    year = "2003"
}

@article{Hu:2012wx,
    author = "Hu, Yuting and Wan, Yidun and Wu, Yong-Shi",
    title = "{Twisted quantum double model of topological phases in two dimensions}",
    eprint = "1211.3695",
    archivePrefix = "arXiv",
    primaryClass = "cond-mat.str-el",
    doi = "10.1103/PhysRevB.87.125114",
    journal = "Phys. Rev. B",
    volume = "87",
    number = "12",
    pages = "125114",
    year = "2013"
}

@article{Wan:2014woa,
    author = "Wan, Yidun and Wang, Juven C. and He, Huan",
    title = "{Twisted Gauge Theory Model of Topological Phases in Three Dimensions}",
    eprint = "1409.3216",
    archivePrefix = "arXiv",
    primaryClass = "cond-mat.str-el",
    doi = "10.1103/PhysRevB.92.045101",
    journal = "Phys. Rev. B",
    volume = "92",
    number = "4",
    pages = "045101",
    year = "2015"
}

@article{Gaiotto:2014kfa,
    author = "Gaiotto, Davide and Kapustin, Anton and Seiberg, Nathan and Willett, Brian",
    title = "{Generalized Global Symmetries}",
    eprint = "1412.5148",
    archivePrefix = "arXiv",
    primaryClass = "hep-th",
    doi = "10.1007/JHEP02(2015)172",
    journal = "JHEP",
    volume = "02",
    pages = "172",
    year = "2015"
}

@article{Schafer-Nameki:2023jdn,
    author = "Schafer-Nameki, Sakura",
    title = "{ICTP lectures on (non-)invertible generalized symmetries}",
    eprint = "2305.18296",
    archivePrefix = "arXiv",
    primaryClass = "hep-th",
    doi = "10.1016/j.physrep.2024.01.007",
    journal = "Phys. Rept.",
    volume = "1063",
    pages = "1--55",
    year = "2024"
}

@article{Bhardwaj:2023kri,
    author = "Bhardwaj, Lakshya and Bottini, Lea E. and Fraser-Taliente, Ludovic and Gladden, Liam and Gould, Dewi S. W. and Platschorre, Arthur and Tillim, Hannah",
    title = "{Lectures on generalized symmetries}",
    eprint = "2307.07547",
    archivePrefix = "arXiv",
    primaryClass = "hep-th",
    doi = "10.1016/j.physrep.2023.11.002",
    journal = "Phys. Rept.",
    volume = "1051",
    pages = "1--87",
    year = "2024"
}

@article{Chatterjee:2022tyg,
    author = "Chatterjee, Arkya and Wen, Xiao-Gang",
    title = "{Holographic theory for continuous phase transitions: Emergence and symmetry protection of gaplessness}",
    eprint = "2205.06244",
    archivePrefix = "arXiv",
    primaryClass = "cond-mat.str-el",
    doi = "10.1103/PhysRevB.108.075105",
    journal = "Phys. Rev. B",
    volume = "108",
    number = "7",
    pages = "075105",
    year = "2023"
}

@article{Chatterjee:2022kxb,
    author = "Chatterjee, Arkya and Wen, Xiao-Gang",
    title = "{Symmetry as a shadow of topological order and a derivation of topological holographic principle}",
    eprint = "2203.03596",
    archivePrefix = "arXiv",
    primaryClass = "cond-mat.str-el",
    doi = "10.1103/PhysRevB.107.155136",
    journal = "Phys. Rev. B",
    volume = "107",
    number = "15",
    pages = "155136",
    year = "2023"
}

@article{Putrov:2016qdo,
    author = "Putrov, Pavel and Wang, Juven and Yau, Shing-Tung",
    title = "{Braiding Statistics and Link Invariants of Bosonic/Fermionic Topological Quantum Matter in 2+1 and 3+1 dimensions}",
    eprint = "1612.09298",
    archivePrefix = "arXiv",
    primaryClass = "cond-mat.str-el",
    doi = "10.1016/j.aop.2017.06.019",
    journal = "Annals Phys.",
    volume = "384",
    pages = "254--287",
    year = "2017"
}

@article{He:2016xpi,
    author = "He, Huan and Zheng, Yunqin and von Keyserlingk, Curt",
    title = "{Field theories for gauged symmetry-protected topological phases: Non-Abelian anyons with Abelian gauge group $\mathbb Z_2^{\otimes 3}$}",
    eprint = "1608.05393",
    archivePrefix = "arXiv",
    primaryClass = "cond-mat.str-el",
    doi = "10.1103/PhysRevB.95.035131",
    journal = "Phys. Rev. B",
    volume = "95",
    number = "3",
    pages = "035131",
    year = "2017"
}

@article{Kaidi:2023maf,
    author = "Kaidi, Justin and Nardoni, Emily and Zafrir, Gabi and Zheng, Yunqin",
    title = "{Symmetry TFTs and anomalies of non-invertible symmetries}",
    eprint = "2301.07112",
    archivePrefix = "arXiv",
    primaryClass = "hep-th",
    doi = "10.1007/JHEP10(2023)053",
    journal = "JHEP",
    volume = "10",
    pages = "053",
    year = "2023"
}

@article{Roumpedakis:2022aik,
    author = "Roumpedakis, Konstantinos and Seifnashri, Sahand and Shao, Shu-Heng",
    title = "{Higher Gauging and Non-invertible Condensation Defects}",
    eprint = "2204.02407",
    archivePrefix = "arXiv",
    primaryClass = "hep-th",
    reportNumber = "YITP-SB-2022-14",
    doi = "10.1007/s00220-023-04706-9",
    journal = "Commun. Math. Phys.",
    volume = "401",
    number = "3",
    pages = "3043--3107",
    year = "2023"
}

@article{Bergman:2026lnz,
    author = {Bergman, Oren and Heckman, Jonathan J. and H{\"u}bner, Max and Migliorati, Daniele and Yu, Xingyang and Zhang, Hao Y.},
    title = "{On the SymTFTs of Finite Non-Abelian Symmetries}",
    journal="{}",
    eprint = "2603.12323",
    archivePrefix = "arXiv",
    primaryClass = "hep-th",
    month = "3",
    year = "2026"
}

@article{Kapustin:2013uxa,
    author = "Kapustin, Anton and Thorngren, Ryan",
    title = "{Higher Symmetry and Gapped Phases of Gauge Theories}",
    eprint = "1309.4721",
    archivePrefix = "arXiv",
    primaryClass = "hep-th",
    doi = "10.1007/978-3-319-59939-7_5",
    journal = "Prog. Math.",
    volume = "324",
    pages = "177--202",
    year = "2017"
}

@article{PhysRevB.87.155114,
  title = {Symmetry protected topological orders and the group cohomology of their symmetry group},
  author = {Chen, Xie and Gu, Zheng-Cheng and Liu, Zheng-Xin and Wen, Xiao-Gang},
  journal = {Phys. Rev. B},
  volume = {87},
  issue = {15},
  pages = {155114},
  numpages = {48},
  year = {2013},
  month = {Apr},
  publisher = {American Physical Society},
  doi = {10.1103/PhysRevB.87.155114},
  url = {https://link.aps.org/doi/10.1103/PhysRevB.87.155114}
}

@article{PhysRevB.71.045110,
  title = {String-net condensation: A physical mechanism for topological phases},
  author = {Levin, Michael A. and Wen, Xiao-Gang},
  journal = {Phys. Rev. B},
  volume = {71},
  issue = {4},
  pages = {045110},
  numpages = {21},
  year = {2005},
  month = {Jan},
  publisher = {American Physical Society},
  doi = {10.1103/PhysRevB.71.045110},
  url = {https://link.aps.org/doi/10.1103/PhysRevB.71.045110}
}

@article{doi:10.1142/S0217979290000139,
author = {WEN, X. G.},
title = {TOPOLOGICAL ORDERS IN RIGID STATES},
journal = {International Journal of Modern Physics B},
volume = {04},
number = {02},
pages = {239-271},
year = {1990},
doi = {10.1142/S0217979290000139},
URL = {
        https://doi.org/10.1142/S0217979290000139
},
eprint = { 
        https://doi.org/10.1142/S0217979290000139
}
}

@article{PhysRevB.82.155138,
  title = {Local unitary transformation, long-range quantum entanglement, wave function renormalization, and topological order},
  author = {Chen, Xie and Gu, Zheng-Cheng and Wen, Xiao-Gang},
  journal = {Phys. Rev. B},
  volume = {82},
  issue = {15},
  pages = {155138},
  numpages = {28},
  year = {2010},
  month = {Oct},
  publisher = {American Physical Society},
  doi = {10.1103/PhysRevB.82.155138},
  url = {https://link.aps.org/doi/10.1103/PhysRevB.82.155138}
}

@article{PhysRevX.8.021074,
  title = {Classification of $\mathbf{(}3+1\mathbf{)}\mathrm{D}$ Bosonic Topological Orders: The Case When Pointlike Excitations Are All Bosons},
  author = {Lan, Tian and Kong, Liang and Wen, Xiao-Gang},
  journal = {Phys. Rev. X},
  volume = {8},
  issue = {2},
  pages = {021074},
  numpages = {24},
  year = {2018},
  month = {Jun},
  publisher = {American Physical Society},
  doi = {10.1103/PhysRevX.8.021074},
  url = {https://link.aps.org/doi/10.1103/PhysRevX.8.021074}
}

@article{Heidenreich:2021xpr,
    author = "Heidenreich, Ben and McNamara, Jacob and Montero, Miguel and Reece, Matthew and Rudelius, Tom and Valenzuela, Irene",
    title = "{Non-invertible global symmetries and completeness of the spectrum}",
    eprint = "2104.07036",
    archivePrefix = "arXiv",
    primaryClass = "hep-th",
    reportNumber = "ACFI-T21-03",
    doi = "10.1007/JHEP09(2021)203",
    journal = "JHEP",
    volume = "09",
    pages = "203",
    year = "2021"
}

@article{Benini:2018reh,
    author = "Benini, Francesco and C{\'o}rdova, Clay and Hsin, Po-Shen",
    title = "{On 2-Group Global Symmetries and their Anomalies}",
    eprint = "1803.09336",
    archivePrefix = "arXiv",
    primaryClass = "hep-th",
    reportNumber = "SISSA 10/2018/FISI, SISSA-10-2018-FISI",
    doi = "10.1007/JHEP03(2019)118",
    journal = "JHEP",
    volume = "03",
    pages = "118",
    year = "2019"
}

@article{Robbins:2025puq,
    author = "Robbins, Daniel and Roy, Subham",
    title = "{SymTFT actions, Condensable algebras and Categorical anomaly resolutions}",
    eprint = "2509.05408",
    archivePrefix = "arXiv",
    primaryClass = "hep-th",
    month = "9",
    year = "2025",
    journal=""
}

@article{Yu:2023nyn,
    author = "Yu, Xingyang",
    title = "{Noninvertible symmetries in 2D from type IIB string theory}",
    eprint = "2310.15339",
    archivePrefix = "arXiv",
    primaryClass = "hep-th",
    doi = "10.1103/PhysRevD.110.065008",
    journal = "Phys. Rev. D",
    volume = "110",
    number = "6",
    pages = "065008",
    year = "2024"
}

@article{Franco:2024mxa,
    author = "Franco, Sebastian and Yu, Xingyang",
    title = "{Generalized symmetries in 2D from string theory: SymTFTs, intrinsic relativeness, and anomalies of non-invertible symmetries}",
    eprint = "2404.19761",
    archivePrefix = "arXiv",
    primaryClass = "hep-th",
    doi = "10.1007/JHEP11(2024)004",
    journal = "JHEP",
    volume = "11",
    pages = "004",
    year = "2024"
}

@article{Bergman:2024its,
    author = "Bergman, Oren and Mignosa, Francesco",
    title = "{String theory and the SymTFT of 3d orthosymplectic Chern-Simons theory}",
    eprint = "2412.00184",
    archivePrefix = "arXiv",
    primaryClass = "hep-th",
    doi = "10.1007/JHEP04(2025)047",
    journal = "JHEP",
    volume = "04",
    pages = "047",
    year = "2025"
}

@article{Heckman:2022xgu,
    author = "Heckman, Jonathan J. and Hubner, Max and Torres, Ethan and Yu, Xingyang and Zhang, Hao Y.",
    title = "{Top down approach to topological duality defects}",
    eprint = "2212.09743",
    archivePrefix = "arXiv",
    primaryClass = "hep-th",
    doi = "10.1103/PhysRevD.108.046015",
    journal = "Phys. Rev. D",
    volume = "108",
    number = "4",
    pages = "046015",
    year = "2023"
}

@article{BANKS198990,
title = {Effective lagrangian description on discrete gauge symmetries},
journal = {Nuclear Physics B},
volume = {323},
number = {1},
pages = {90-94},
year = {1989},
issn = {0550-3213},
doi = {https://doi.org/10.1016/0550-3213(89)90589-0},
url = {https://www.sciencedirect.com/science/article/pii/0550321389905890},
author = {Thomas Banks}
}

@article{wilzeck,
  title = {Discrete gauge symmetry in continuum theories},
  author = {Krauss, Lawrence M. and Wilczek, Frank},
  journal = {Phys. Rev. Lett.},
  volume = {62},
  issue = {11},
  pages = {1221--1223},
  numpages = {0},
  year = {1989},
  month = {Mar},
  publisher = {American Physical Society},
  doi = {10.1103/PhysRevLett.62.1221},
  url = {https://link.aps.org/doi/10.1103/PhysRevLett.62.1221}
}

@article{PRESKILL199050,
title = {Local discrete symmetry and quantum-mechanical hair},
journal = {Nuclear Physics B},
volume = {341},
number = {1},
pages = {50-100},
year = {1990},
issn = {0550-3213},
doi = {https://doi.org/10.1016/0550-3213(90)90262-C},
url = {https://www.sciencedirect.com/science/article/pii/055032139090262C},
author = {John Preskill and Lawrence M. Krauss}
}

@article{Gao:2025ihw,
    author = "Gao, Zhi-Qiang and Liu, Chunxiao and Moore, Joel E.",
    title = "{Topological BF Theory construction of twisted dihedral quantum double phases from spontaneous symmetry breaking}",
    eprint = "2511.19589",
    archivePrefix = "arXiv",
    primaryClass = "cond-mat.str-el",
    month = "11",
    journal = {arXiv preprint},
    year = "2025"
}
\end{document}